\documentclass[12pt]{article}
\usepackage[utf8]{inputenc}
\usepackage[dvips, includefoot, margin=1in]{geometry}
\usepackage{setspace}
\usepackage{bm}
\usepackage{amsmath, amssymb, amsthm, amsfonts, amssymb}
\usepackage{enumitem}
\usepackage{color}
\usepackage{hhline}
\usepackage{float}
\RequirePackage{graphicx}
\usepackage[caption=false]{subfig}
\usepackage[authoryear]{natbib}
\usepackage{authblk}
\usepackage{multicol}
\usepackage{multirow}
\RequirePackage[colorlinks,citecolor=blue,urlcolor=blue]{hyperref}

\title{\bf Latent trajectory models for spatio-temporal dynamics in Alaskan ecosystems}
\author[1]{Xinyi Lu\thanks{Corresponding author. Email: xinyi.lu@colostate.edu.}}
\author[2]{Mevin B. Hooten}
\author[3, 4]{Ann M. Raiho}
\author[5]{David K. Swanson}
\author[6, 7]{Carl A. Roland}
\author[6, 7]{Sarah E. Stehn}
\affil[1]{\small\textit{Department of Statistics, Colorado State University, Fort Collins, Colorado 80523 USA}}
\affil[2]{\small\textit{Department of Statistics and Data Sciences, The University of Texas at Austin, Austin, Texas 78712 USA}}
\affil[3]{\small\textit{NASA Goddard Space Flight Center, Greenbelt, Maryland 20771 USA}}
\affil[4]{\small\textit{University of Maryland Earth System Science Interdisciplinary Center, College Park, Maryland 20740 USA}}
\affil[5]{\small\textit{National Park Service, 4175 Geist Road, Fairbanks, Alaska 99709 USA}}
\affil[6]{\small\textit{Denali National Park and Preserve, P.O. Box 9, Denali Park, Alaska 99755 USA}}
\affil[7]{\small\textit{Central Alaska Network Inventory and Monitoring Program, 4175 Geist Road, Fairbanks, Alaska 99709 USA}}
\date{}

\begin{document}

\maketitle

\section*{Abstract}
The Alaskan landscape has undergone substantial changes in recent decades, most notably the expansion of shrubs and trees across the Arctic. We developed a dynamic statistical model to quantify the impact of climate change on the structural transformation of ecosystems using remotely sensed imagery. We used latent trajectory processes in a hierarchical framework to model dynamic state probabilities that evolve annually, from which we derived transition probabilities between ecotypes. Our latent trajectory model accommodates temporal irregularity in survey intervals and uses spatio-temporally heterogeneous climate drivers to infer rates of land cover transitions. We characterized multi-scale spatial correlation induced by plot and subplot arrangement in our study system. We also developed a P\'{o}lya-Gamma sampling strategy to improve computation. Our model facilitates inference on the response of ecosystems to shifts in the climate and can be used to predict future land cover transitions under various climate scenarios.
\bigbreak
Keywords: Bayesian; climate change; data augmentation; ecological succession; state-space models.

\section{Introduction}
Climate change can impact ecosystems by altering vegetation composition. A prominent example is the expansion of shrubs, coined ``shrubification,'' due to warming climates in northern Alaska \citep{Swanson2013, Brodie2019}. The encroachment of woody plants reduces erosion \citep{Tape2010}, increases fire frequency \citep{Higuera2008}, decreases albedo, and furthers warming \citep{Chapin2005}. Remotely sensed imagery provides readily available high-resolution information about land cover and are commonly used to understand landscape changes \citep{Svenningsen2015}. In this study, we analysed pairs of historic and contemporary aerial images acquired across central Alaska. Our multi-scale multivariate spatio-temporal model provides inference about the rates of climate-driven land cover transitions among 5 major ecotypes. 
\bigbreak
Discrete-time Markov Chains (DTMCs) characterize a sequence of state changes by their transition probabilities and are commonly used to model landscape change \citep{Baker1989}, community dynamics \citep{Hill2004}, and plant succession \citep{Logofet2000}. In our study, the state space is finite and specified by ecotypes, and state changes occur in discrete time because of distinct growth seasons in Alaska. However, transition probabilities of a DTMC depend directly on the observed states. In the presence of sampling irregularity, modeling transition probabilities by a DTMC would require imputation of ``missing" states at a temporal resolution typically defined by the smallest common sampling interval. Our objective is to learn about transition probabilities at the climate scale of 30 years, whereas the imagery pairs were collected 25 to 32 years apart. A naive DTMC would impute intermediate states annually to account for sampling discrepancy (suggesting that over 90\% of the states need to be imputed). Without regularization on the transition mechanism or techniques such as multiple imputation \citep{Scharf2017, Scharf2019} there is no guarantee that the sequence of states inferred by the DTMC is representative of the progressive changes in plant communities.  
\bigbreak
The evolutionary mechanism of plant succession often sustain dependence beyond the temporal resolution of a DTMC. Higher-order Markov chains and semi-Markov models provide additional flexibility in temporal dynamics but dependence may remain at the state level \citep{Moore1990, Lazrak2010}. We characterize state changes using latent spatio-temporal processes in a logit-transformed probability space. We perceive such latent processes as an ecosystem analogue to animal movement \citep{Mcclintock2014, Hooten2017} and refer to the spatio-temporally evolving state probabilities as the ``latent trajectories" of the ecosystems. Ecosystem trajectories are commonly used in ecology and environmental science to describe the ecosystem dynamics over time \citep{Locatelli2017, Lamothe2019}, and in our method such trajectories are represented by their ordination in the latent probability space. Our latent trajectory representation redirects dependence between states to that between latent locations in the transformed probability space, thereby circumventing imputation of intermediate states. Our approach is related to spatial process models for non-Gaussian data via generalized linear modeling \citep{Diggle1998, Finley2009}. \cite{Jin2013} and \cite{Berrett2016} respectively demonstrated the utility of spatial generalized linear (mixed) models in land cover classification. When temporal dynamics are involved, \cite{Bradley2019} showed that multinomial spatio-temporal mixed effects models can be used to analyze high-dimensional longitudinal data. Modeling latent processes in a continuous space provides additional smoothness to modeling observed processes in a discrete space and leads to consistent inference on ecotypes from year to year. Different trajectory models reflect various evolution mechanisms of ecosystems we are able to accommodate. For example, the development of a primary forest may be represented by a random-walk-with-drift model directing at high probability regions of the dominant community. Successional impact of invasive species may be represented by interactions between the different dimensions of a latent trajectory, and extreme weather events that trigger secondary succession may be incorporated as Dirac delta functions in the latent trajectory. We illustrate these concepts in detail in Section \ref{latent_traj}.
\bigbreak
Further, our model achieves sampling efficiency through P\'{o}lya-Gamma data augmentation \citep{Polson2013}. Similar to the Albert-Chib data augmentation for probit regression \citep{AlbertChib1993}, sampling auxiliary P\'{o}lya-Gamma random variables in a logistic regression model promotes conjugacy of linear predictors. We extend the P\'{o}lya-Gamma approach to multinomial logistic regression and incorporate multi-scale spatial correlation for the case study.

\section{Imagery Data}
\begin{figure}[H]
    \centering
    \includegraphics[scale=0.6]{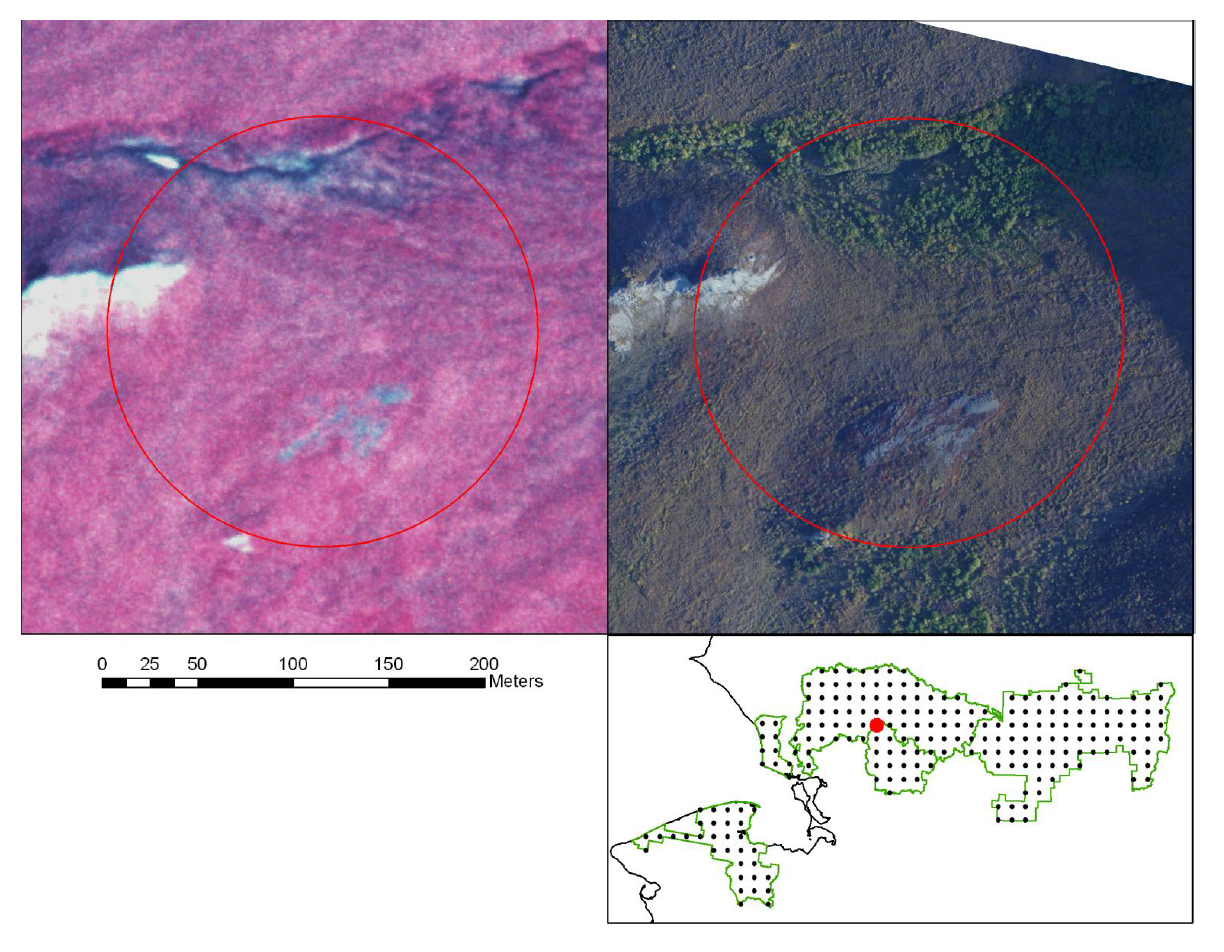}
    \caption{A pair of images collected in 1979 and 2008, respectively, in the Noatak National Preserve, Alaska, as represented by the red dot on the inset map showing all plot locations for the case study. The historic image (AHAP color-infrared photo, left) consists mostly of herbaceous and low shrub vegetation. The contemporary image (small-format true color photo, right) consists mostly of tall shrubs. Source: Swanson, 2013. }
    \label{image_pair}
\end{figure}

\begin{figure}[H]
    \centering
    \includegraphics[scale=0.8, trim=10mm 0mm 10mm 0mm]{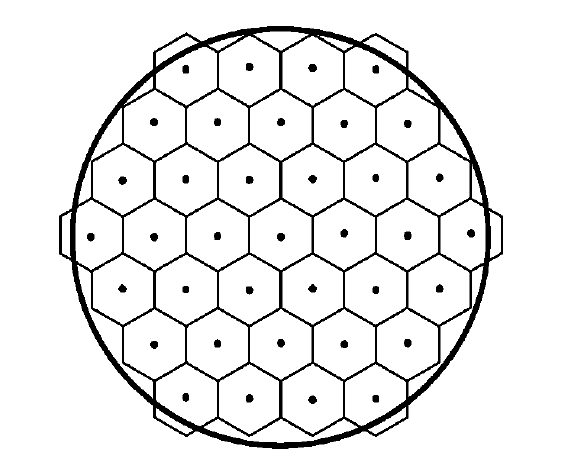}
    \caption{A four-hectare sample plot with 37 contiguous regular hexagonal subplots for ecotype classification. Source: Swanson, 2013.}
    \label{subplots}
\end{figure}

The data motivating our analysis comprise 200 aerial imagery pairs acquired across the NPS Arctic Inventory and Monitoring Network (ARCN) from 1977 to 2010. A pair of aerial images consists of a georeferenced high-resolution color digital aerial image taken between 2008-2010 and a scanned and georeferenced color-infrared aerial image taken between 1977-1985 of the same plot on a systematic grid over ARCN (Figure \ref{image_pair}). Each image was further divided into 37 contiguous regular hexagonal subplots during processing (Figure \ref{subplots}) and their dominant ecotypes were visually determined according to the scheme developed for the ARCN Ecological Land Survey and Land Cover Map \citep{Jorgenson2009}. We do not account for classification error in this study because a single highly-experienced observer processed all imagery pairs \citep{Swanson2013}. We categorized the 44 ecotypes developed for the Land Cover Map by their vegetation biomass and composition into the following 5 major ecotypes:
\begin{itemize}
    \item \textbf{Forest}: Lowland Black Spruce Forest, Riverine Poplar Forest, Riverine White Spruce-Poplar Forest, Riverine White Spruce-Willow Forest, Upland Birch Forest, Upland Spruce-Birch Forest, Upland White Spruce Forest, Upland White Spruce-Lichen Woodland;
    \item \textbf{Tall Shrub}: Lowland Alder Tall Shrub, Riverine Alder or Willow Tall Shrub, Upland Alder-Willow Tall Shrub;
    \item \textbf{Low Shrub}: Alpine Dryas Dwarf Shrub, Alpine Ericaceous Dwarf Shrub, Coastal Crowberry Dwarf Shrub, Lowland Birch-Ericaceous-Willow Low Shrub, Lowland Ericaceous Shrub Bog, Lowland Willow Low Shrub, Riverine Birch-Willow Low Shrub, Riverine Dryas Dwarf Shrub, Riverine Willow Low Shrub, Upland Birch-Ericaceous-Willow Low Shrub, Upland Dwarf Birch-Tussock Shrub, Upland Willow Low Shrub;
    \item \textbf{Barren}: Alpine Acidic Barrens, Alpine Acidic Barrens, Alpine Mafic Barrens,  Coastal Barrens, Human Modified Barrens, Riverine Barrens, Upland Mafic Barrens, Upland Sandy Barrens;
    \item \textbf{Other}: Alpine Lake, Alpine Wet Sedge Meadow, Coastal Brackish Sedge–Grass Meadow, Coastal Dunegrass Meadow, Coastal Water, Lowland Lake, Lowland Sedge-Dryas Meadow, Lowland Sedge Fen, Riverine Water, Riverine Wet Sedge Meadow, Shadow/ Indeterminate, Snow, Upland Sedge-Dryas Meadow.
\end{itemize}
The above more generalized ecotype categorization accounts for 90\% of the total transitions between the original 44 mapped ecotypes. The 10\% of transitions unaccounted for are intra-category. Table \ref{sum_trans_type} summarizes the observed transition frequencies by subplots. Most of the subplots experienced no change over the study period. Among the 5 major ecotypes used for our analysis, the most frequent transitions were from Low Shrub to Forest, Low Shrub to Tall Shrub, and Other to Barren. Empirical studies suggested that most Shrub to Forest transitions occurred by post-fire succession, and most Low Shrub to Tall Shrub transitions occurred by tundra shrub increase; most transitions between Barren and Other occurred by fluvial processes in riverine environments, and a few other transitions occurred by thermokarst \citep{Swanson2013}.  

\section{Model}
\subsection{Data Model}
We let $\bm{y}_{i, s, t}$ be a $K$-dimensional standard unit vector with one denoting the observed state (ecotype) of plot $i$, subplot $s$, at time $t$. For $i = 1, \dots, n_I$ (number of plots), $s = 1, \dots, n_S$ (number of subplots per plot), and $t \in \mathbb{N}$ (years), we model $\bm{y}_{i, s, t}$ as 
\begin{align}
    \bm{y}_{i, s, t} &\sim \text{MN}\left(1, \bm{p}_{i, s, t}\right),\\
    \bm{p}_{i, s, t} &= \pi_{\text{SB}}^{-1}(\tilde{\bm{p}}_{i, s, t}),
\end{align}
where $\pi_{SB}(\cdot): [0, 1]^K \rightarrow [0, 1]^{K-1}$ is a bijective mapping with
\begin{align*}
    \tilde{p}_{i, s, t, 1} &= p_{i, s, t, 1},\\
    \tilde{p}_{i, s, t, k} &= \frac{p_{i, s, t, k}}{1 - \sum_{r < k}p_{i, s, t, r}},\ k = 2, \dots, K - 1.
\end{align*}
The mapping is known as a stick-breaking transformation similar to the stick-breaking process used in the construction of Dirichlet processes \citep{Ishwaran2001}. The $k$th element of $\tilde{\bm{p}}_{i, s, t}$ represents the conditional probability that $\bm{y}_{i, s, t}$ is in state $k$ given it is not in any of the states $1, \dots, k-1$. This allows us to express the probability mass function of $\bm{y}_{i, s, t}$ as a product of conditional binomials,
\begin{equation}\label{sb}
    \left[\bm{y}_{i, s, t}\middle|\tilde{\bm{p}}_{i, s, t}\right] = \prod_{k = 1}^{K-1}\text{Binom}(y_{i, s, t, k};N_{i, s, t, k}, \tilde{p}_{i, s, t, k}),
\end{equation}
where $N_{i, s, t, 1} = 1$ and $N_{i, s, t, k} = 1 - \sum_{r<k}y_{i, s, t, r}$ for $k = 2, \dots, K-1$. The stick-breaking transformation allows us to exploit conjugacy in each of the binomial models through P\'{o}lya-Gamma data augmentation \citep{Linderman2015}, and we describe this strategy in Section \ref{PG}. Although the stick-breaking process implies a prior stochastic ordering, it is not of practical concern when the model is data driven and the state space finite. In addition, the stick-breaking representation of multinomial logistic regression can be more economical than the alternative representation in \cite{Holmes2006} because it does not require evaluating a proportionality constant for every state-specific parameter.

\subsection{Latent Trajectory Model}\label{latent_traj}
\begin{figure}[H]
    \centering
    \includegraphics[scale=0.6, trim=0mm 10mm 0mm 20mm]{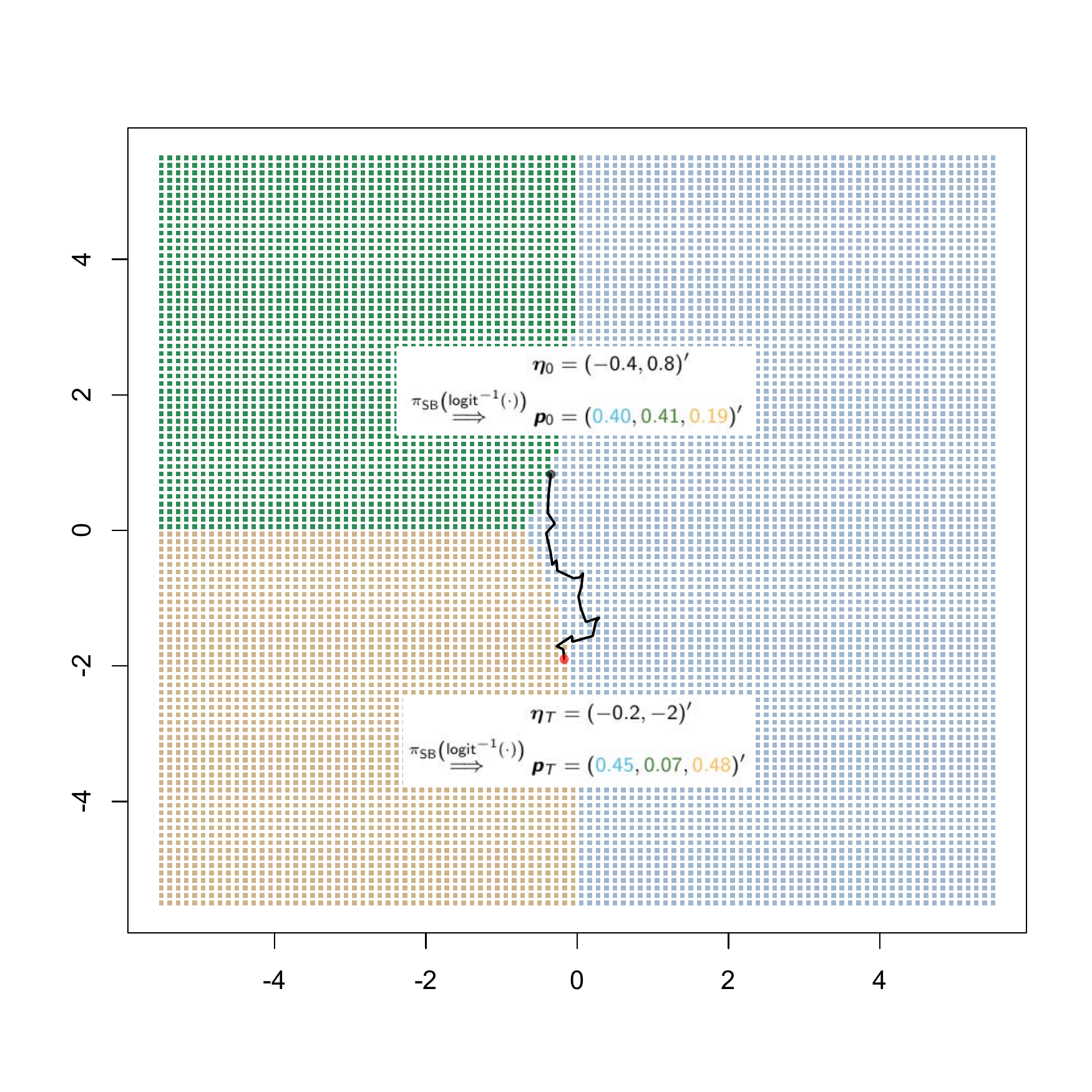}
    \caption{A simulated trajectory in a two-dimensional logit-transformed probability space. Each location is colored by the most probable state. A simulated trajectory shows changes in state probabilities as an ecosystem travels across the latent space.}
    \label{rw_eg}
\end{figure}

We specify a logit function that maps $\tilde{\bm{p}}_{i, s, t}$ to $\bm{\eta}_{i, s, t}$ for plot $i$, subplot $s$, at time $t$ in the $(K-1)$-dimensional real space as $\text{logit}\left(\tilde{\bm{p}}_{i, s, t}\right) = \bm{\eta}_{i, s, t}$. We model the dynamics in $\tilde{\bm{p}}_{i, s, t}$ through a trajectory in the logit-transformed probability space ($\bm{\eta}$-space). For illustration, Figure \ref{rw_eg} shows a two-dimensional $\bm{\eta}$-space where each location is translated to a three-dimensional vector representing its probabilities in Shrub (blue), Forest (green), and Barren (yellow) states. The locations are colored by their most probable states. The simulated ecosystem trajectory starts at $\bm{\eta}_0 = (-0.4, 0.8)'$ (black point), which corresponds to state probabilities $\bm{p}_0 = (0.40, 0.41, 0.19)'$, indicating that the ecosystem is likely to be Shrub or Forest and unlikely to be Barren. At the end of the trajectory (red point), the ecosystem is at $\bm{\eta}_T = (-0.2, -2)'$, which corresponds to state probabilities $\bm{p}_T = (0.45, 0.07, 0.48)'$, indicating that the ecosystem is likely to be Shrub or Barren and unlikely to be Forest after time $T$. An ecosystem in the $\bm{\eta}$-space will always have non-zero probability in all states; however, state probabilities can become highly concentrated as the ecosystem departs from the origin (e.g., at $\bm{\eta}_t = (7, 0)'$, the associated state probabilities, $\bm{p}_t$, are almost $(1, 0, 0)'$).
\bigbreak
Among an array of discrete-time continuous-space trajectory models, we specify a random-walk-with-drift model because it is a simple model that accommodates a temporal trend. The drift in the latent trajectory indicates the direction of ecological succession. For example, post-fire colonization of forests is represented by a drift vector directing from regions with high barren  probabilities to regions with high forest probabilities, and the magnitude of the drift implies the rate of succession. We have, for $k = 1, \dots, K - 1$,
$$
   \eta_{i, s, t, k} = \eta_{i, s, t - 1, k} + \delta_{i, s, t, k} + e_{i, s, t, k}, 
$$
where $\delta_{i, s, t, k}$ is the spatio-temporally varying drift and $e_{i, s, t, k}$ represents uncertainty in the trajectory. The latent location of the ecosystem at time $T$ is therefore
\begin{align*}
       \eta_{i, s, T, k} &= \eta_{i, s, 0, k} +  \sum_{t = 1}^T\left(\eta_{i, s, t, k} - \eta_{i, s, t-1, k}\right), \\
       &= \eta_{i, s, 0, k} + \sum_{t = 1}^T\delta_{i, s, t, k} + \sum_{t = 1}^Te_{i, s, t, k}.
\end{align*}
We model the initial conditions, $\eta_{i, s, 0, k}$, with landscape covariates, $\bm{h}_{i, s}$, as follows
\begin{equation}\label{eta0}
    \eta_{i, s, 0, k} = \bm{h}_{i, s}'\bm{\alpha}_k + \zeta_{i, s, k},
\end{equation}
where the Gaussian random effects, $\zeta_{i, s, k} \sim \text{N}\left(0, \sigma^2_\zeta\right)$, provide additional flexibility. We model the drifts, $\delta_{i, s, t, k}$, using climate covariates, $\bm{x}_{i, s, t}$, as follows
\begin{equation}\label{delta}
    \delta_{i, s, t, k} = \bm{x}_{i, s, t}'\bm{\beta}_k.
\end{equation}
We further decompose movement uncertainty, $e_{i, s, t, k}$, into two-levels of spatial random effects. At the plot level, we account for correlation among locations on a systematic grid (Figure \ref{image_pair}), $\bm{\xi}_{t, k}$, using a geostatistical model. At the subplot level, we account for correlation among the contiguous regular hexagons within the same plot (Figure \ref{subplots}), $\bm{\epsilon}_{i, t, k}$, using an intrinsic conditional autoregressive (ICAR) model, so that for $t = 1, \dots, T$ and $k = 1, \dots, K-1$,
\begin{align}
      \bm{\xi}_{t, k} &\sim \text{N}\left(\bm{0}, \sigma^2_\xi\exp(-\bm{D}/\phi)\right),\\
      \bm{\epsilon}_{i, t, k} &\sim \text{N}\left(\bm{0}, \sigma^2_\epsilon(\bm{R} - \bm{W})^{-1}\right),
\end{align}
where $\bm{D}$ represents geodesic distances (in kilometers) between plots, $\bm{W}$ is the adjacency matrix for subplots within a plot, and $\bm{R}$ is the diagonal matrix with row sums of $\bm{W}$ as diagonal elements \citep{VerHoef2018}. The ecosystem trajectory at plot $i$, subplot $s$, in the $k$th dimension can be summarized as
\begin{align}\label{eta}
    \eta_{i, s, T, k} &= \bm{h}_{i, s}'\bm{\alpha}_k + \zeta_{i, s, k} + \sum_{t = 1}^{T}\left(\bm{x}_{i, s, t}'\bm{\beta}_k + \xi_{i, t, k} + \epsilon_{i, s, t, k}\right), \nonumber\\
    &= \bm{h}_{i, s}'\bm{\alpha}_k + \zeta_{i, s, k} +\underbrace{\bm{\beta}_k'\sum_{t = 1}^T\bm{x}_{i, s, t} + \sum_{t = 1}^T\xi_{i, t, k} + \sum_{t = 1}^T\epsilon_{i, s, t, k}}_{\Delta_{i, s, T, k}}.
\end{align}
The image pairs in our study have different beginning and ending years, and the dependence structure of the latent trajectories need to account for overlapping time intervals at different plots. The spatio-temporal covariance at the plot level is therefore 
\begin{align}\label{cov_xi}
    \text{Cov}\left(\xi_{i, t_1 k}, \xi_{j, t_2, k}\right) &= \left\{\begin{array}{l}
    \sigma^2_\xi,\ \text{if $i = j$ and $t_1 = t_2$};\\
    \sigma^2_\xi\exp(-D_{ij}/\phi),\ \text{if $i \neq j$ and $t_1 = t_2$};\\
    0,\ \text{if $t_1 \neq t_2$}.
    \end{array}\right.
\end{align}
The spatio-temporal covariance at the subplot level (within the same plot $i$) is 
\begin{align}\label{cov_eps}
\text{Cov}\left(\epsilon_{i, s, t_1, k}, \epsilon_{i, r, t_2, k}\right) &= \left\{\begin{array}{l}
\sigma^2_\epsilon,\ \text{if $s = r$ and $t_1 = t_2$};\\
\sigma^2_\epsilon Q_{ij},\ \text{if $s \neq r$ and $t_1 = t_2$};\\
0,\ \text{if $t_1 \neq t_2$},
\end{array}\right.
\end{align}
where $\bm{Q} = (\bm{R} - \bm{W})^{-1}$. To maintain propriety in the ICAR model, we constrain the subplot level random effects to sum to zero by sampling $\epsilon_{i, s, t, k}$ via conditioning by kriging \citep{Rue2005}.
\bigbreak
We specify exchangeable Gaussian priors for the covariate coefficients, $\bm{\alpha}_k$ and $\bm{\beta}_k$, for $k = 1, \dots, K-1$, and we specify Inverse-Gamma priors for the variance parameters, $\sigma^2_\zeta$, $\sigma^2_\epsilon$, and $\sigma^2_\xi$. The range parameter $\phi$ is given a uniform prior bounded above by 1/3 of the maximum distance between plots. We provide a full description of the priors used for the case study in Web Appendix B.

\subsection{P\'{o}lya-Gamma Data Augmentation}\label{PG}
Every binomial component of the multinomial likelihood in (\ref{sb}) can be expressed as  $y_{i, s, t, k} \sim \text{Binom}(N_{i, s, t, k}, \text{logit}^{-1}(\eta_{i, s, t, k}))$ for $i = 1, \dots, n$. By Theorem 1 in \cite{Polson2013}, the following integral identity holds for the binomial likelihood $[y_{i, s, t, k}|\eta_{i, s, t, k}]$,
\begin{equation}\label{theorem1}
\frac{\{\exp(\eta_{i, s, t, k})\}^{y_{i, s, t, k}}}{\{1 + \exp(\eta_{i, s, t, k})\}^{N_{i, s, t, k}}} = 2^{-{N_{i, s, t, k}}}\exp(\kappa_i\eta_i)\int_0^\infty\exp\left(-\omega\eta_i^2/2\right)p(\omega)d\omega,
\end{equation}
where $\kappa_{i, s, t, k} = y_{i, s, t, k} - N_{i, s, t, k}/2$ and $\omega\sim\text{PG}(N_{i,s , t, k}, 0)$. Because the right hand side of (\ref{theorem1}) contains a Gaussian kernel when conditioned on the P\'{o}lya-Gamma random variable $\omega$, normal conjugacy holds for $\bm{\eta}_{i, s, t}$ under a normal prior, $\bm{\eta}_{i, s, t} \sim \text{N}\left(\bm{\mu}_{i, s, t}, \bm{\Sigma}_{i, s, t}\right)$. The posterior distribution of $\bm{\eta}_{i, s, t}$ is 
$$
\left[\bm{\eta}_{i, s, t}\middle|\bm{y}_{i, s, t}, \bm{\omega}_{i, s, t}\right] = \text{N}\left(\bm{m}_{i, s, t}, \bm{V}_{i, s, t}\right),
$$
where
\begin{align*}
\bm{V}_{i, s, t} &= \left(\bm{\Sigma}^{-1}_{i, s, t} + \bm{\Omega}_{i, s, t}\right)^{-1},\ \bm{m}_{i, s, t} = \bm{V}_{i, s, t}\left(\bm{\Sigma}^{-1}_{i, s, t}\bm{\mu}_{i, s, t} + \bm{\kappa}_{i, s, t}\right),\\
\bm{\Omega}_{i, s, t} &= \text{diag}(\bm{\omega}_{i, s, t}),\ \bm{\kappa}_{i, s, t} = \bm{y}_{i, s, t} - \bm{N}_{i, s, t}/2.
\end{align*}
Conjugacy also holds for the P\'{o}lya-Gamma random variables. The posterior distribution of $\omega_{i, s, t, k}$ is
$$
\left[\omega_{i, s, t, k}\middle|\eta_{i, s, t, k}\right] = \text{PG}\left(N_{i, s, t, k}, \eta_{i, s, t, k}\right),
$$
for $k = 1, \dots, K-1$. The P\'{o}lya-Gamma approach aligns well with our stick-breaking representation of the multinomial likelihood and facilitates conjugacy in the linear predictors and the spatial random effects of the latent trajectory model. Although \citeauthor{Johndrow2018} (\citeyear{Johndrow2018}) suggested that Metropolis-Hastings algorithms may be more efficient than data augmentation schemes including P\'{o}lya-Gamma with imbalanced categorical data, empirical examination of the traceplots in our study did not suggest significant autocorrelation. The data augmentation approach circumvents tuning, and is therefore particularly helpful to implement complex models such as ours. We provide a detailed description of the MCMC algorithm developed for this study in Web Appendix A.

\subsection{Transition Matrix}
Transition matrices can be used to compare our model inference to the observed transition frequencies and illustrate land cover changes under future climate scenarios. Our model allows us to infer instantaneous state probabilities from latent locations in the $\bm{\eta}$-space, and we obtain elements of the transition matrix as a derived quantity. Because the transition mechanism of latent trajectory models differs entirely from that of DTMCs, the derived transition matrices do not possess Markovian properties such as phase-type distributions and stationary distributions. Nonetheless, our model answers relevant ecological questions such as species persistence and climax community using posterior predictive realizations of states. A posterior predictive realization of the transition matrix over time $T$ given the $q$th posterior sample, $\bm{\eta}_{i, s, 0}^{(q)}$ and $\bm{\Delta}_{i, s, T}^{(q)}$ (see (\ref{eta}) for definition), for $q = 1, \dots, Q$, is derived as follows,
\begin{enumerate}
    \item Sample a posterior predictive realization of the initial state as 
    $$
    \bm{y}_{i, s, 0}^{(q)}\sim \text{MN}\left(1, \pi_{\text{SB}}^{-1}\left(\text{logit}^{-1}\left(\bm{\eta}_{i, s, 0}^{(q)}\right)\right)\right)
    $$ 
    for $i = 1, \dots, n_I$ and $s = 1, \dots, n_S$;
    \item Sample a posterior predictive realization of the state at time $T$ as 
    $$
    \bm{y}_{i, s, T}^{(q)}\sim \text{MN}\left(1, \pi_{\text{SB}}^{-1}\left(\text{logit}^{-1}\left(\bm{\eta}_{i, s, 0}^{(q)} + \bm{\Delta}_{i, s, T}^{(q)}\right)\right)\right)
    $$ 
    for $i = 1, \dots, n_I$ and $s = 1, \dots, n_S$;
    \item Calculate a posterior predictive realization of the transition probability from state $k_0$ to $k_1$ over time $T$, for $k_0 = 1, \dots, K$ and $k_1 = 1, \dots, K$, as 
    $$
    M_{k_0, k_1}^{(q)} = \frac{\sum_{i = 1}^{n_I}\sum_{s = 1}^{n_S}\mathbb{I}\left(y_{i, s, 0, k_0} = 1, y_{i, s, T, k_1} = 1\right)}{\sum_{i = 1}^{n_I}\sum_{s = 1}^{n_S}\mathbb{I}\left(y_{i, s, 0, k_0} = 1\right)},
    $$ 
    where $\bm{M}^{(q)}$ is the $q$th posterior predictive transition matrix.
\end{enumerate}
The posterior mean predictive transition matrix is evaluated as $\widehat{\bm{M}} = \frac{1}{Q}\sum_{q = 1}^Q\bm{M}^{(q)}$, and element-wise credible intervals can be constructed by applying quantile functions to the $Q$ realizations of $\bm{M}$.
\bigbreak
We use simulation to illustrate that our model is able to recover the covariate coefficients and spatial parameters in Web Appendix C. 

\section{Case Study}
\begin{figure}[H]
    \centering
    \includegraphics{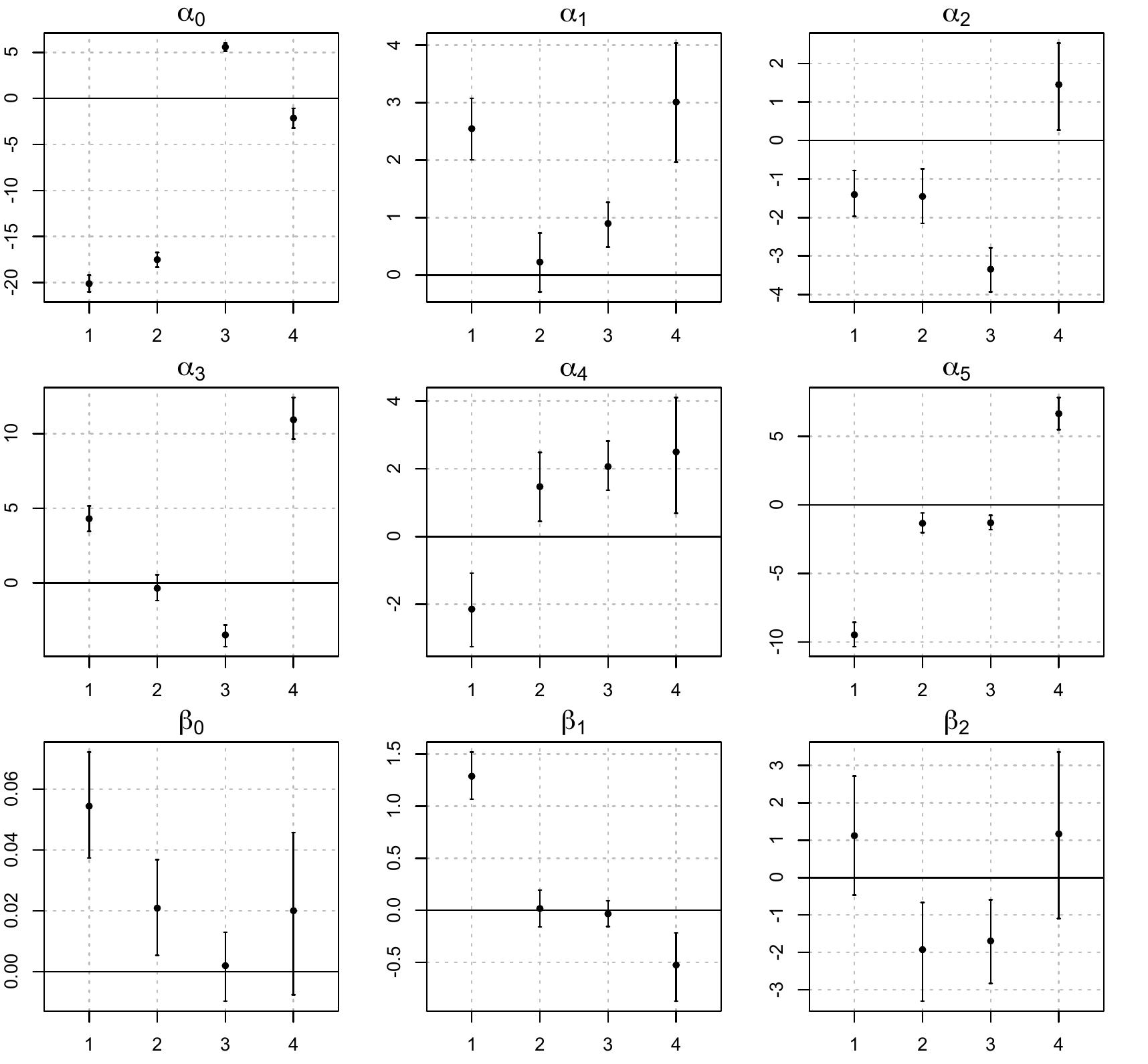}
    \caption{Estimated posterior means (black circles) and the corresponding 95\% credible intervals (line segments) for the case study. The x-axes represent the ecotypes (i.e., the second indices) associated with each explanatory variable. The y-axes represent the parameter values.}
    \label{cov_est}
\end{figure}

\begin{figure}[H]
    \centering
    \subfloat[]{\label{trans_mat_est}\includegraphics[width=0.5\textwidth, trim=0mm 30mm 0mm 20mm]{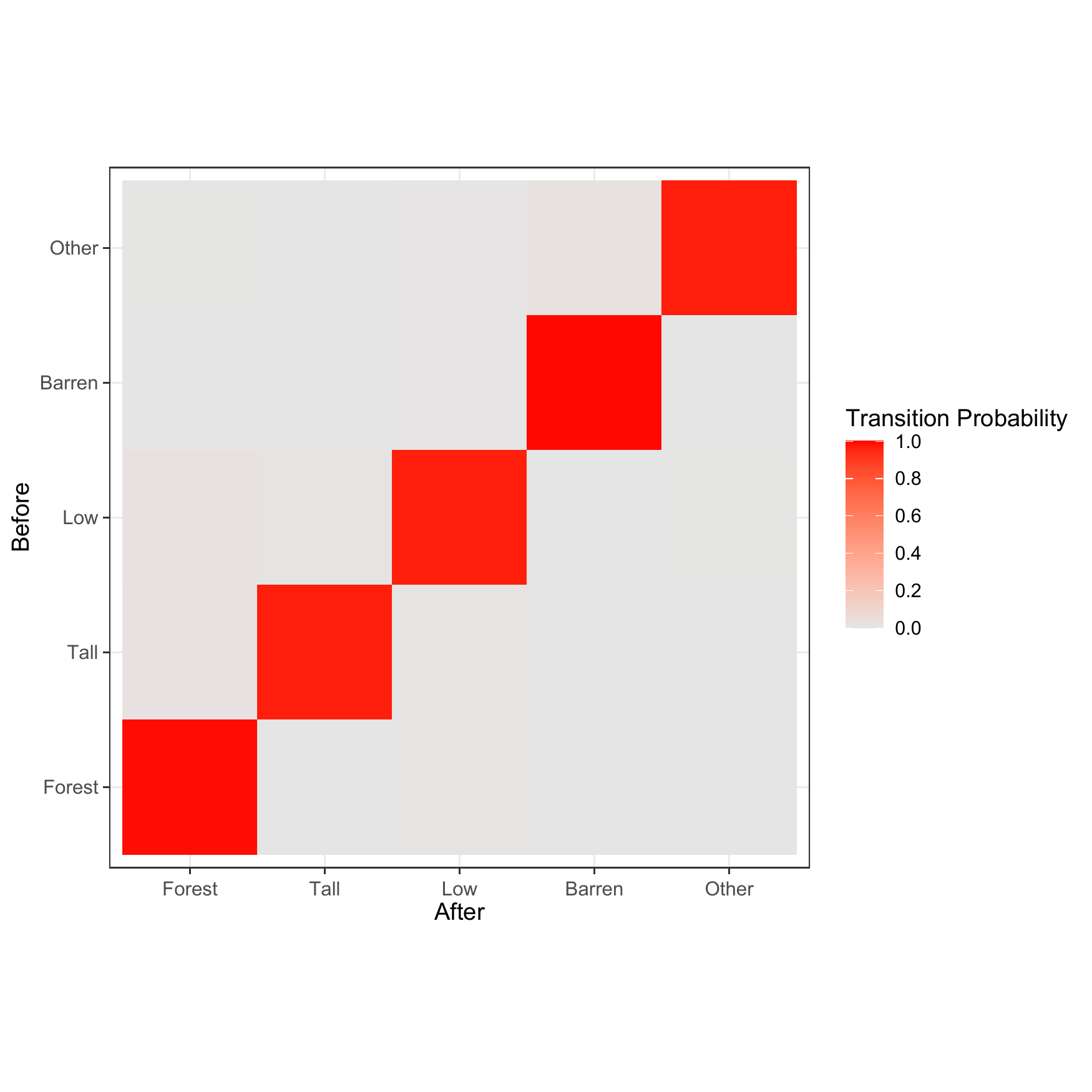}}
    \subfloat[]{\label{trans_mat_est_od}\includegraphics[width=0.5\textwidth, trim=0mm 30mm 0mm 20mm]{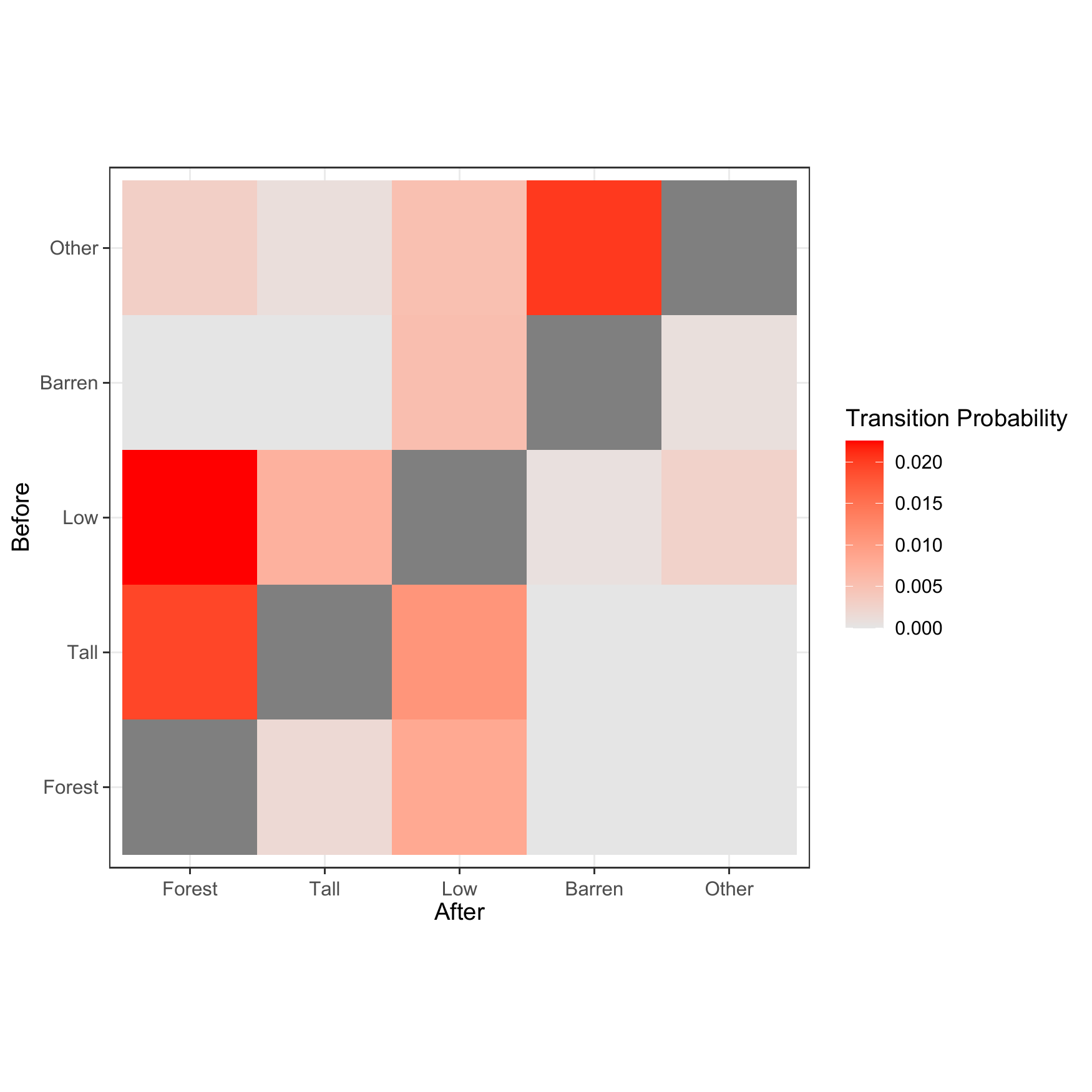}}\\
    \caption{Illustrations of (a) the posterior mean predictive transition matrix and (b) the off-diagonal elements of the posterior mean predictive transition matrix for the case study. Rows indicate categorization in c. 1980 and columns indicate categorization in c. 2010.}
\end{figure}

\begin{figure}[H]
    \centering
    \subfloat[]{\label{trans_mat_hi}\includegraphics[width=0.5\textwidth, trim=0mm 30mm 0mm 20mm]{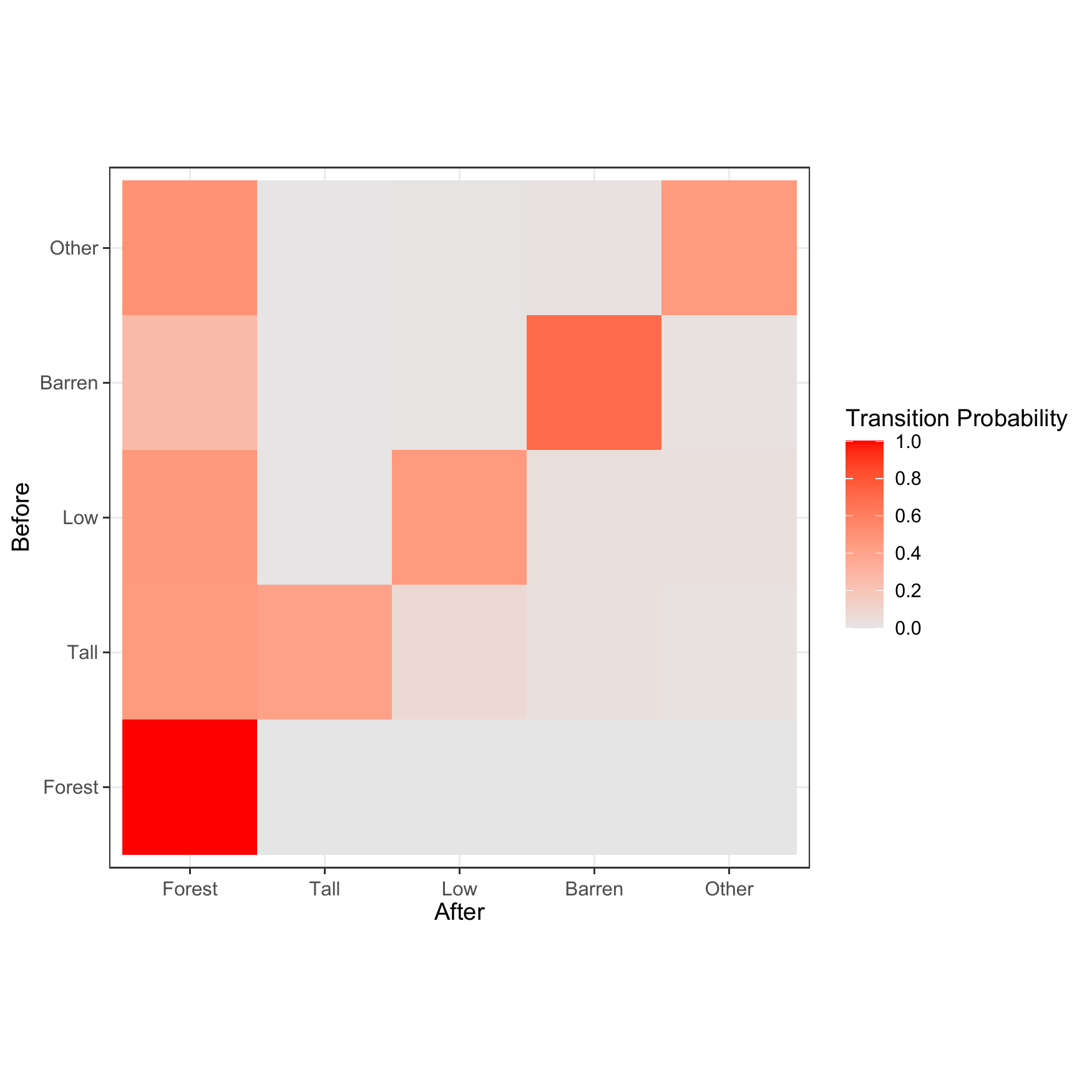}}
    \subfloat[]{\label{trans_mat_lo}\includegraphics[width=0.5\textwidth, trim=0mm 30mm 0mm 20mm]{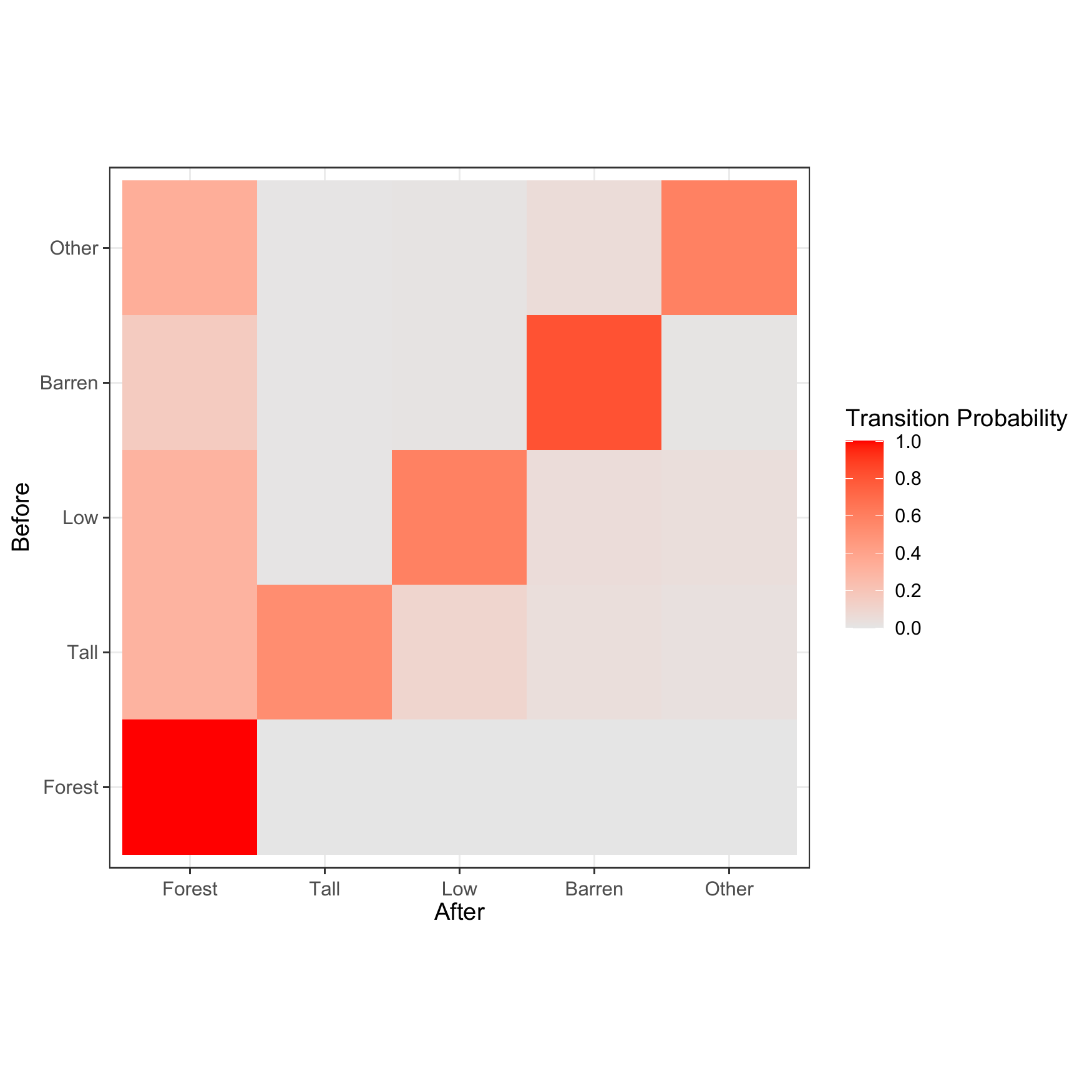}}\\
    \caption{The posterior mean predictive transition matrices under (a) a high emission scenario (time = 120, temp = 8, pcpt = 2); and (b) a low emission scenario (time = 120, temp = 4, pcpt = 2).}
\end{figure}

The landscape and the climate variables only vary at the plot level in our application due to their large-scale spatial resolutions. Summer temperature (July in Northern Hemisphere) and soil moisture are the two major drivers of tundra vegetation biomass and composition \citep{Murray1982, Elmendorf2012}. For the initial condition model in (\ref{eta0}), we used the covariate vector, $\bm{h}_{i, s}$, that includes an intercept, mean July temperature in the first year (Celcius), aspect degree (azimuth clockwise from north), slope degree, interaction between slope and aspect, and elevation (feet). The interaction term quantifies potential insolation in addition to what is accounted for by temperature. We mapped aspect onto a linear spectrum using Beers' transformation \citep{Beers1966} and standardized all covariates besides the intercept by subtracting their means and dividing by their standard deviations. For the drift model in (\ref{delta}), we used the covariate vector, $\bm{x}_{i, s, t}$,  that includes an intercept, change in mean July temperature from year $t-1$ to $t$ (Celcius), and change in mean daily precipitation from year $t-1$ to $t$ (mm). We obtained daily temperature and precipitation from the downscaled European Centre for Medium-Range Weather Forecasts Re-Analysis (ERA)-Interim historical reanalysis data (SNAP) at 20km spatial resolution over the state of Alaska. Because these environmental data were unavailable prior to 1979, we used the 1979 data as a proxy for images collected in 1977 and 1978 (78 out of 200). We aggregated daily measurements at each plot to obtain their mean July temperature and mean daily precipitation in the corresponding years. We ran the MCMC algorithm in R version 3.0.2 \citep{team2019r} for 10,000 iterations and used a burn-in of 2,000 iterations. Our algorithm took 2 hours on a 2.5GHz Intel Core i5 processor. Figure \ref{cov_est} illustrates the posterior distributions of the covariate coefficients. Table 3 (Web Appendix D) summarizes the posterior estimates and the convergence statistics of the covariate coefficients and the spatial parameters in our case study.
\bigbreak
Each covariate coefficient is first indexed by the explanatory variable and then by the ecotype (Forest = 1, Tall Shrub = 2, Low Shrub = 3, Barren = 4, and Other is the reference category). Due to the stick-breaking transformation, the $k$th element of a coefficient vector represents the change in \textit{conditional} log odds with a unit increment in the corresponding covariate given that a subplot is not in any of the previous states $1, \dots, k-1$. The $\alpha$ coefficients explain the association between the landscape covariates and the distribution of ecotypes in c. 1980. Inference on the intercept vector ($\bm{\alpha}_0$) agrees with the empirical frequencies in c. 1980 (i.e., row sums in Table \ref{sum_trans_type}). The negative estimated intercepts for Forest ($\alpha_{01}$) and Tall Shrub ($\alpha_{02}$) indicate low initial probabilities in these states, and the positive estimated intercept for Low Shrub ($\alpha_{03}$) indicate a high initial probability in the state. We estimated a positive temperature coefficient ($\alpha_{11}$) and a negative elevation coefficient ($\alpha_{51}$) for Forest, suggesting that forest ecotypes are likely to occupy plots at low elevations with warm growing seasons. The model fit also suggested that barren ecotypes are likely to occupy plots at high elevations ($\alpha_{54}$) with steep slopes ($\alpha_{34}$).
\bigbreak
The $\beta$ coefficients explain the temporal dynamics of, and the effect of climate change on, land cover state probabilities. The time coefficients for Forest ($\beta_{01}$) and Tall Shrub ($\beta_{02}$) have positive estimated 95\% credible intervals, suggesting that probability mass will accumulate in these states over time. The coefficients for Low Shrub ($\beta_{03}$) and Barren ($\beta_{04}$) both have zero-overlapping estimated 95\% credible intervals; however, as the probabilities in Forest and Tall Shrub grow, probabilities in other states will likely decline. We estimated a positive temperature coefficient for Forest ($\beta_{11}$) and a negative temperature coefficient for Barren ($\beta_{14}$), suggesting that warmer climates will lead to more frequent forest ecotypes and cooler climates will lead to more frequent barren ecotypes. We estimated negative precipitation coefficients for Tall Shrub ($\beta_{22}$) and Low Shrub ($\beta_{23}$), suggesting that drier climates are conducive to more frequent shrub ecotypes. We obtained posterior predictive realizations of transition probabilities following Section 2.5 for a variety of climate scenarios. Although inference on the covariate coefficients depends on the order of ecotypes, the derived transition matrices demonstrate consistent patterns regardless of ecotype ordering. Figure \ref{trans_mat_est} shows the posterior mean predictive transition matrix for the case study, which is validated by the empirical transition frequencies in Table \ref{sum_trans_type}. 
\bigbreak
Temperature and precipitation are highly variable across Alaska. Nonetheless, the state has experienced an overall warming more than twice as fast as the contiguous U.S. in recent decades, with the most dramatic changes in spring and winter \citep{Stewart2013}. Studies project that the annual mean temperature will increase from 4 to 10 degrees Celsius by the end of this century under higher emission scenarios or from 2 to 6 degrees Celsius under lower emission scenarios \citep{Stewart2017}. The annual precipitation in Alaska is also projected to increase by 10\% or more by mid-century \citep{Stewart2017}. We projected land cover transitions under the assumption of uniform climate change over the study area for the purpose of demonstration. Our model inference can be used in conjunction with spatially detailed climate forecasts to obtain more realistic predictions on landscape transformation.  
\bigbreak
Figure \ref{trans_mat_hi} shows the posterior mean predictive transition matrix under a high emission scenario where we assume that July temperature increases by 8 degrees Celsius and daily precipitation increases by 2mm (equivalent to a 730mm annual increment) uniformly in the study region from c. 1980 to 2100. As suggested by the $\bm{\beta}_0$ estimates, probabilities will accumulate in Forest from all other states in c. 120 years. Most predicted transitions into Forest are from Other, Tall Shrub, and Low Shrub, possibly due to succession facilitated by warming climates. In comparison, there are fewer transitions from Barren to Forest, possibly due to landscape factors (e.g., temperature and elevation) that limit forest expansion. There are also few transition from Low Shrub to Tall Shrub, possibly because low shrubs undergoing succession have passed the state of Tall Shrub and reached the state of Forest after over a century. Figure \ref{trans_mat_lo} shows the posterior mean predictive transition matrix under a low emission scenario where we assume that July temperature increases by 4 degrees Celsius and daily precipitation increases by 2mm uniformly in the study region from c. 1980 to 2100. Table 4 (Web Appendix D) summarizes the posterior predictive mean transition probabilities and their associated uncertainty under the high and the low emission scenarios, respectively. The posterior mean transition probabilities are generally lower under the low emission scenario than those under the high emission scenario, although the differences between the two scenarios are not great. Our predicted transition patterns agree with the observed patterns in Figure \ref{trans_mat_est} where the most frequent transitions are those from low biomass categories to high biomass categories. The transition probabilities are magnified under hypothetical warming scenarios and serve to support the association between Arctic greening and climate change on the global scale \citep{Epstein2004, Harsch2009}. Lastly, our predictions of landscape transitions do not account for uncertainty in climate forecasts and the variability in our predictions increases with the length of projection into the future.

\section{Discussion}
We presented a latent trajectory model for landscape change while accounting for spatio-temporal dependence. Our model characterizes dependence in the logit-transformed probability space and leverages computational efficiency through P\'{o}lya-Gamma data augmentation. We demonstrated dependence structures that manifest evolutionary mechanisms; therefore, our model is most suitable when the transition process takes place at a lower temporal frequency than that of the data collection. The latent trajectory models provide insight into the cumulative effect of long-term warming on Alaskan landscape and allow inference about state distributions and transition probabilities over flexible time intervals. 
\bigbreak
We decomposed the variance in the latent trajectory process into three sources: the uncertainty in the initial conditions, $\sigma^2_\zeta$, the plot level uncertainty, $\sigma^2_\xi$, and the subplot level uncertainty, $\sigma^2_\epsilon$. The three parameters are identifiable because $\sigma^2_\zeta$ is informed by both the historic and the contemporary images, whereas $\sigma^2_\xi$ and $\sigma^2_\epsilon$ are informed by the contemporary images and modeled at different spatial scales. The estimated $\sigma^2_\zeta$ is larger than the other two variance parameters (Table \ref{cov_est}) because the effects of $\sigma^2_\xi$ and $\sigma^2_\epsilon$ are scaled by time. The large initial variance may be attributed to few observations in the temporal domain and  scarcity of transitions in our case study. The uncertainty in the initial conditions could hinder prediction at unobserved locations and potentially confound with the latent movement processes. We regularize the estimation of initial conditions by incorporating spatial structure and leveraging the fact that the observations at different plots were staggered in time. Further, the uncertainty may be reduced with more temporal replicates at each plot and more diverse transition types. The $\bm{\alpha}$ estimates are substantial in magnitude. They produce initial conditions with highly concentrated probabilities, which demonstrates our learning about the temporal dynamics of the transition process. An uneven initial condition (with a significant probability in one state and negligible probabilities in the other states) indicates that the ecosystem is unlikely to experience any systematic change in the immediate future, possibly due to limiting environmental factors explained by the covariate vector $\bm{h}_i$. On the other hand, the $\bm{\beta}$ estimates representing movement along the latent trajectories are smaller because fundamental changes in ecosystems usually take place over the course of centuries. As such, our model was able to account for small-scale temporal irregularity within the large-scale ecological process.  
\bigbreak
There are several ways to extend our latent trajectory model. A constant time coefficient ($\bm{\beta}_0$) implies that, over time, probability mass will likely converge in one state as an ecosystem moves toward regions with high probabilities in that state. Such behaviors may be appropriate for some types of succession (e.g., primary/secondary successions where the dominant state is stable), but not otherwise (e.g., seasonal successions where the dominant state alternates frequently). A more flexible process model could use higher-order terms, interactions, or basis functions to represent the temporal trend. For succession mechanisms that are characterized by the coexistence or competition between species, we can explicitly model interactions between states through the covariance of the $K-1$ dimensions of $\bm{\eta}$. Lastly, environmental events such as fire and flooding often result in substantial changes in the ecosystem. These events can be important predictors of transformation and incorporating these events into dynamic models is a potential area of future research. In our application, however, only 2 out of the 200 plots had a record of fire prior to the collection of historic images, and a larger spatio-temporal domain is required to capture the effects of fires.
\bigbreak
Our hierarchical framework aids in integrating various data sources, some of which may arise from other studies on Alaskan vegetation change. For example, \cite{Scharf2022} and \cite{Raiho2022} developed a model to identify landscape factors that resist climate-driven vegetation change. Although limited by the absence of temporal replicates, \cite{Raiho2022} made inference over a greater spatial domain and performed statistical model selection to distinguish key variables. Their analysis output --- a measurement of ecosystem robustness to climate change given its landscape covariates --- could be used in our latent trajectory model because ``robust" ecosystems should move less in the $\bm{\eta}$-space due to climate change. On the other hand, our model is based on temporal replicates, and we could compare our inference with the results from \cite{Raiho2022} in a future study to visualize how our estimated climate effects relate to the robustness scores.

\section*{Acknowledgments}
This research was funded by the National Park Service and NSF DEB 1927177. We thank numerous technicians for assistance in the field and lab.

\newpage
\bibliographystyle{biom.bst}
\bibliography{refs}

\section*{Supporting Information}
Web Appendices and Tables referenced in Sections 3 and 4 are available with this paper at the Biometrics website on Wiley Online Library.

\newpage
\begin{table}[H]
    \centering
    \caption{Summary of transition frequencies by subplots. Rows indicate categorization in c. 1980 and columns indicate categorization in c. 2010.}
    \begin{tabular}{c c c c c c c c}\hline
      & \multicolumn{7}{c}{(c. 2010)} \\
      & & Forest & Tall & Low & Barren & Other & Sum \\\hline
      \multirow{5}{*}{\rotatebox{90}{(c. 1980)}} & Other & 4 & 4 & 17 & 26 & 944 & 995\\
      & Barren & 0 & 0 & 17 & 1107 & 4 & 1128\\
      & Low & 117 & 66 & 4043 & 0 & 4 & 4230\\
      & Tall & 15 & 440 & 0 & 0 & 0 & 455\\
      & Forest & 587 & 0 & 0 & 0 & 5 & 592\\
      & Sum & 723 & 510 & 4077 & 1133 & 957 & 7400\\\hline
    \end{tabular}
    \label{sum_trans_type}
\end{table}

\end{document}